\documentclass[12pt,a4paper]{article}
\usepackage{amsthm,amsfonts,amsmath,amssymb,array,graphicx,tabularx}
\usepackage[mathscr]{eucal}

\usepackage[paperwidth=20.5cm,paperheight=29.7cm,width=18cm,height=25.5cm,top=24mm,left=14mm,headsep=11mm]{geometry}

\def\supind#1{${}^\mathrm{#1}$}
\def\title#1{\begin{center}{\LARGE\bf #1}\end{center}}
\renewcommand{\author}[2]{\begin{center}{#1}\\ \medskip{\small\it #2}\end{center}}
\renewcommand{\epsilon}{\varepsilon}
\renewcommand{\kappa}{\varkappa}

\newcommand{\rot}{\mathop{\rm rot}\nolimits}
\newcommand{\Div}{\mathop{\rm div}\nolimits}
\newcommand{\grad}{\mathop{\rm grad}\nolimits}
\newcommand{\erf}{\mathop{\rm erf}\nolimits}
\renewcommand{\Im}{\mathop{\rm Im}\nolimits}
\renewcommand{\Re}{\mathop{\rm Re}\nolimits}

\newcommand{\ex}{\mathop{\rm e}\nolimits}
\newcommand{\exi}[1]{\mathop{\rm e}\nolimits^{{\mathrm i}#1}}

\newcommand{\irm}{\mathrm i}
\newcommand{\Ls}{\mathcal L}

\begin{document}

\title{Electromagnetic fields induced by surface ring\\ \medskip
 waves in the deep sea}

\author{Kozitskiy~S.B.\supind{1}}
{\supind{1}Il'ichev Pacific Oceanological Institute, 43 Baltiiskaya St., Vladivostok, 690041, Russia\\
e-mail: {\tt skozi@poi.dvo.ru} }

\begin{abstract}
The paper deals with electromagnetic effects associated with a
radially symmetric system of progressive surface waves in the deep
sea, induced by underwater oscillating sources or by dispersive
decay of the initial localized perturbations of the sea surface.
Key words: surface waves, electromagnetic field variations,
magnetic hydrodynamics.
\end{abstract}

\section*{Introduction}

In this article we derive formulas describing the variation of the electromagnetic
fields induced by the radially symmetric system of progressive
surface waves on the surface of a conductive liquid.

Motion of a conductive fluid in a constant external magnetic field at
small magnetic Reynolds number, for example, of the sea water,
is accompanied by an interconnected system of electromagnetic fields and
currents, which has almost no reverse effect on the liquid movement itself.
Experimental study of electromagnetic fields with the use of
both contact and remote measurement techniques
provides information on the dynamics and parameters of the original
hydrodynamic process which presents some practical interest.

\section{Electromagnetic fields induced by progressive ring waves}

Obtain analytical solutions for the electromagnetic field variations from
surface ring waves excited by oscillating underwater
sources~\cite{B3} for the case of an infinitely deep fluid with a
constant conductivity throughout the volume, and with the constant external
magnetic field $\vec F$ having $F_z$ vertical and
$F_y$ horizontal components.

Cartesian coordinate system is chosen so that the $z$-axis is
directed vertically upward, and the direction of the $y$-axis coincides with
the direction of the horizontal component of the external magnetic field.
Level of the interface between water and air corresponds to the plain $z = 0$.
In further notation values of electromagnetic fields in the air
will be denoted by a subscript $a$, and the values of the fields and
currents in the water are taken without an index.

The initial equations for determining the electromagnetic quantities
are the Maxwell equations written with known simplifying assumptions \cite{A2}:
\begin{equation}
\begin{split}
& \rot{\vec E} = - \frac{1}{c}\partial_t{\vec B}\,,
\qquad \Div{\vec B} = 0\,, \\
& \rot{\vec B} = \frac{4\pi}{c}{\vec J},
\qquad \Div{\vec D} = 0\,, \\
& \vec J = \sigma{\vec E}+\frac{1}{c}\,[\vec v,\vec F], \qquad
\vec D = \epsilon{\vec E} + \frac{\epsilon -1}{c}\,[\vec v,\vec
F]\,,
\end{split}
\end{equation}
where: $\vec B$ is the magnetic induction, $\vec E$ is the
electric field tension, $\vec D$ is the electric induction, $\vec J$ is
the electric current density, $c$ is the speed of light in vacuum,
$\sigma$ is the fluid conductivity, $\vec v$ is the velocity of the fluid,
$\epsilon$ is the dielectric constant of the medium.

The interface conditions have the following form:
\begin{equation}\label{R2}
D_{na}-D_n = 4\pi q\,, \quad B_{na} = B_n\,, \quad\ {\vec B}_{\tau a}
 = {\vec B}_{\tau}, \quad {\vec E}_{\tau a} = {\vec E}_{\tau}\,.
\end{equation}

Index $n$ denotes the normal to the interface component of the corresponding
vector, $\tau$ denotes the tangential one, $q$ is the surface charge density.
In the case when the fluid velocity field is assumed to be potential,
it is convenient to write these equations and interface conditions through
the magnetic Hertz vector $\vec P$ and the velocity potential $\phi$:
\begin{equation}\label{R3}
\nu_m\Delta\vec P - {\partial_t}\vec P = \phi {\vec F}\,, \qquad
\Delta{\vec P}_a = 0\,.
\end{equation}
Here Hertz vector $\vec P$ along with $\vec F$: $(0, F_y, F_z)$
has two components $\vec P$: $(0, P_y, P_z)$; $\nu_m =
{c^{2}/4\pi\sigma}$ is the magnetic viscosity.

Vector equation (\ref{R3}) can be considered as two scalar equations
for $y$ and $z$ components of the corresponding vectors.
For the air the equation is transformed into the Laplace equation.
The vector components of the field and of current density can be found
by differentiating:
\begin{equation}\label{R4}
\begin{split}
& {\vec B}=\rot\rot{\vec P} = \grad\Div{\vec P}-\Delta{\vec P}\,,  \\
& {\vec E}=-\frac{1}{c}\rot{\partial_t{\vec P}}\,, \\
& {\vec J}=-\frac{\sigma}{c}\rot\left({\partial_t{\vec P}}
+\phi{\vec F}\right) = -\frac{c}{4\pi}\rot\Delta{\vec P}\,,\\
& {\vec v} = - \grad\phi\,.
\end{split}
\end{equation}
Substitution of these expressions in (\ref{R2}) allows to obtain the interface
conditions on the surface $z = 0$ for the magnetic Hertz vector:
\begin{equation}\label{R5}
\begin{split}
& P_z = P_{az}\,, \quad \partial_z P_y = \partial_z P_{ay}\,, \quad
\Delta P_y = \Delta P_{ay}\,,\\
& \partial_z P_z + \partial_y P_y = \partial_z P_{az} + \partial_y
P_{ay}\,.
\end{split}
\end{equation}

Assume that the potential of the fluid velocity satisfies
Laplace equation and has the following form:
\begin{equation}\label{R6}
\phi(x,y,t) = R(x,y)\ex^{kz}\exi{\omega t}\,,
\end{equation}
where the complex function of two variables $R(x,y)$ must satisfy
the Helmholtz equation: $$(\partial^2_x+\partial^2_y)R+k^2 R=0\,.$$
Solutions for the electromagnetic quantities can be obtained without
specifying the form of this function.
Thus the solutions of a class of similar problems, differ in the form of function $R(x,y)$ are determined.
For example, in the case of cylindrical progressive waves propagating from the source,
the velocity potential has the form \cite{B3}:
\begin{equation}\label{R7}
\begin{split}
& \phi(r,t) = A(k)\ex^{kz}(J_0(kr)\sin\omega t-Y_0(kr)\cos\omega t) = \\
& \qquad = \Im( A(k)\ex^{kz}H_0^{(2)}(kr)\exi{\omega t})\,,
\end{split}
\end{equation}
where $r$ is radial coordinate, $H_0^{(2)}$ is Hankel function of the second kind,
$J_0$ and $Y_0$ are Bessel functions.

Amplitude coefficient $A(k)$ depends on the wavenumber $k$ and is
determined by the method of wave excitation. In particular, if the waves
are produced by the pulsing point monopole source located at a depth $h$ and having
performance $Q = Q_0 \cos\omega t$, then $A(k) = (Q_0 k/2)
\ex^{-kh}$. If, for example, the generation of the waves is performed by the
vertical oscillatory movements of the sphere of radius $a$ with an amplitude $M$,
located at a depth of $h$, then $A(k) = M\pi a^3 k^2\sqrt{kg}\ex^{-kh}$ \cite{B3}.

The solution of equations (\ref{R3}) with the potential (\ref{R6})
and with the interface conditions (\ref{R5}) will be sought in the standard
form of a superposition of a particular solution of inhomogeneous and
the general solution of the homogeneous equations:
\begin{equation}\label{anz}
\vec P = -\frac{1}{\irm\omega}\exi{\omega t}{\vec F}R(x,y)\ex^{kz}+
\exi{\omega t}{\vec G}\ex^{\kappa z}\,,{\vec P}_a
= \exi{\omega t}{\vec G}_a\ex^{-kz}\,,
\end{equation}
where the functions $\vec G(x,y)$ and ${\vec G}_a(x,y)$ satisfy
the same Helmholtz equation, as the function $R(x, y)$.
After the imposition the interface conditions (\ref{R5}) on the ansatz (\ref{anz})
on the surface $z=0$ we obtain the following equations for the $\vec G(x, y)$ and ${\vec G}_a(x, y)$:
\begin{equation}
\begin{split}
& - \frac{1}{\irm\omega}F_z R + G_z = G_{az}\,, \qquad
 \frac{1}{\irm\omega}F_y R = G_{ay}\,, \qquad G_y = 0\,, \\
& -\frac{k}{\irm\omega}\left(F_y\frac{\partial}{k\partial y}+F_z\right)R+\kappa G_z
 = -kG_{az}+\frac{\partial G_{ay}}{\partial y} \,.
\end{split}
\end{equation}
Here the solutions for $G_y$ and $G_{ay}$ are ready.
After resolving these expressions with respect to $G_z$ and $G_{az}$ we have:
\begin{equation}
\begin{split}
& G_z = \frac{2k}{\irm\omega(k+\kappa)}\left(F_y\frac{\partial}{k\partial y}+F_z\right)R\,,\\
& G_{az} = \frac{2k}{\irm\omega(k+\kappa)}\left(F_y\frac{\partial}{k\partial y}+F_z\right)R
- \frac{1}{\irm\omega}F_z R\,,\\
\end{split}
\end{equation}
Final solutions for the Hertz vector is written as follows:
\begin{equation}
\begin{split}
& P_y = -\frac{1}{\irm\omega}\exi{\omega t}F_y R(x,y)\ex^{kz}\,,
\qquad  P_{ay} = \frac{1}{\irm\omega}\exi{\omega t}F_y R(x,y)\ex^{-kz}\,,  \\
& P_z = \frac{1}{\irm\omega}\exi{\omega t}
\left[\frac{2k}{\kappa+k}\left(F_y\frac{\partial}{k\partial y}
+F_z\right)\ex^{\kappa z}-F_z\ex^{kz}\right] R(x,y)\,, \\
& P_{az} = \frac{1}{\irm\omega}\exi{\omega
t}\left[\frac{2k}{\kappa+k}\left(F_y\frac{\partial}{k\partial y}
+F_z\right)-F_z\right] R(x,y)\ex^{-kz}\,,
\end{split}
\end{equation}
where the parameter $\kappa$, has the dimension of the wavenumber:
$$
\kappa^2 = k^2 - k_0^2\,, \qquad k_0^2 = -\irm\omega/\nu_m\,.
$$
\begin{figure}[tbh]
\centering{%
\includegraphics[width=0.65\textwidth,angle=-90]{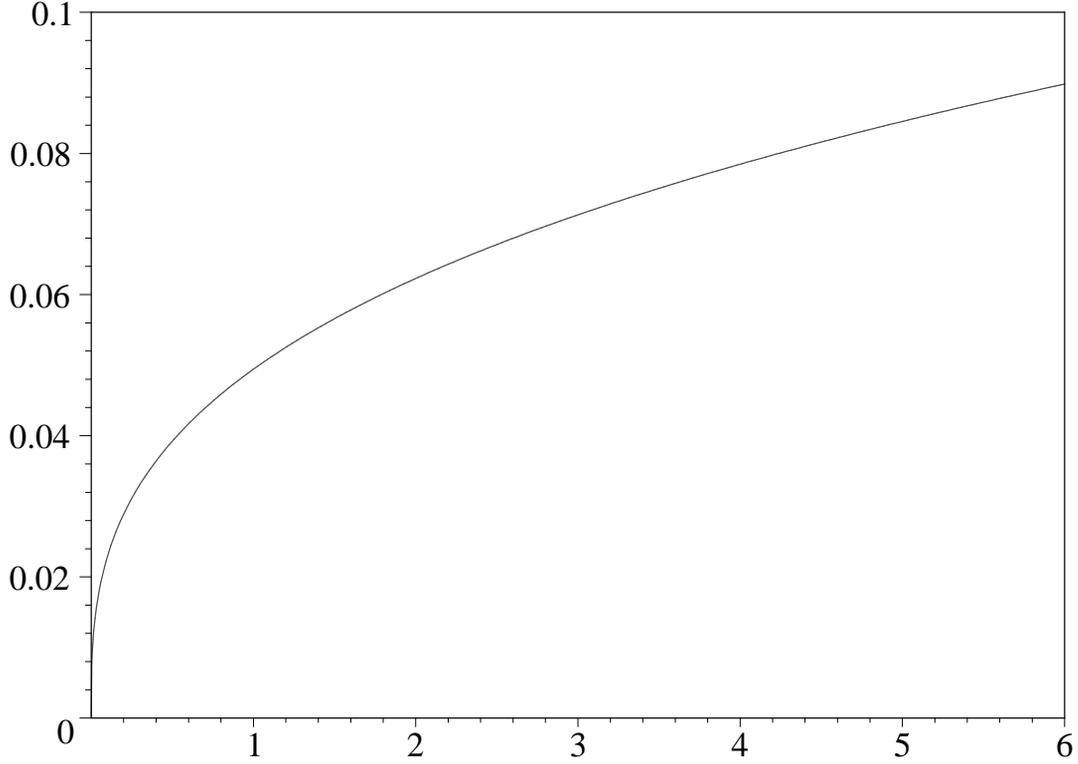}}
\caption{Dependance of $\Omega = (g/c)\sqrt[3]{4\pi\sigma c/g}$ from
$\sigma$ in Sm/m.}
\label{fig1}
\end{figure}

\begin{figure}[tbh]
\centering{%
\includegraphics[width=0.65\textwidth,angle=-90]{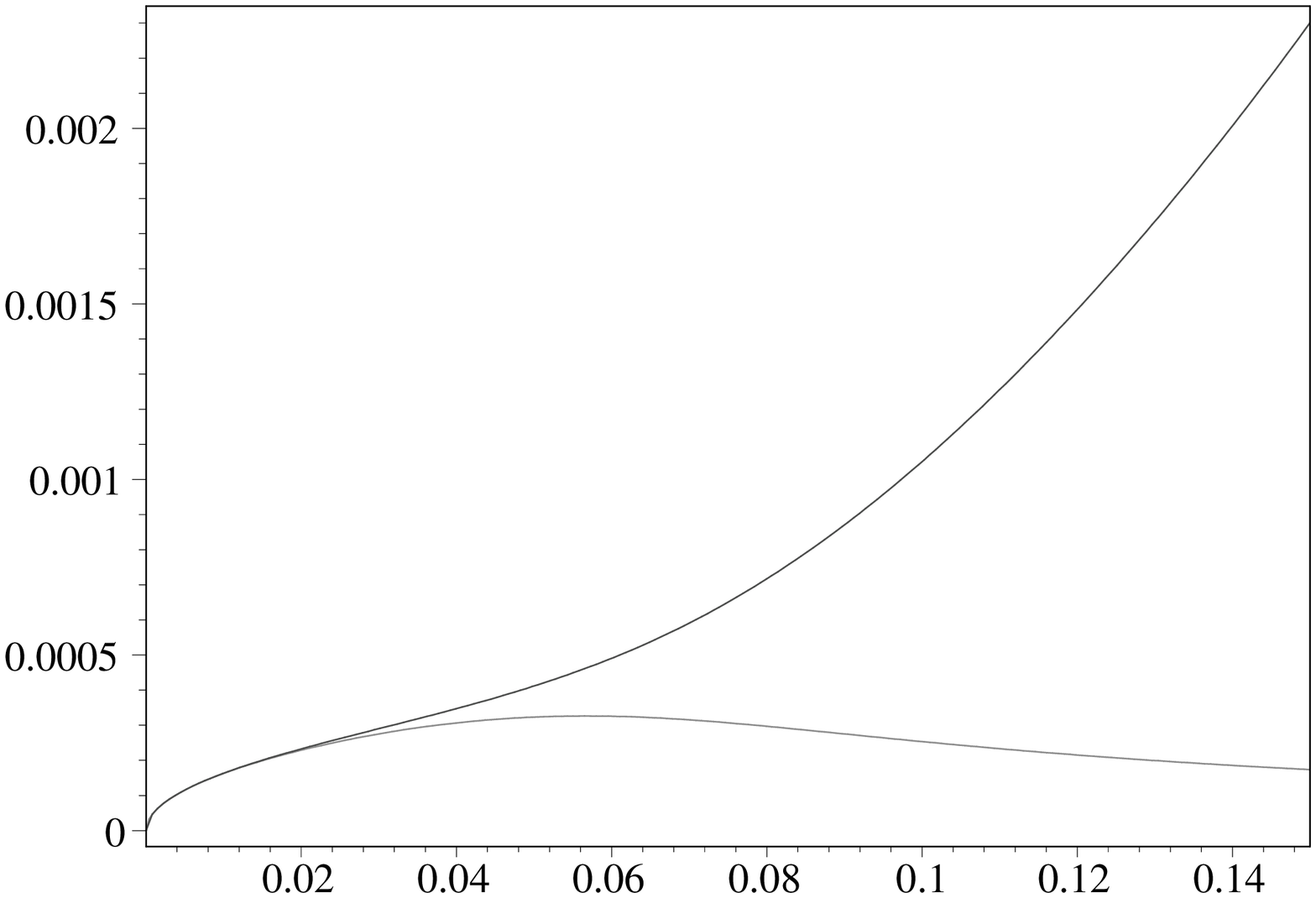}}
\caption{Dependance of $\alpha(\omega)$ (upper curve) and $\beta(\omega)$ (lower curve) from
$\omega$, $\Omega = 0.08$.}
\label{fig2}
\end{figure}

Introduce $\Omega = (g/c)\sqrt[3]{4\pi\sigma c/g}$, where $g$ is the gravity acceleration (see Fig.~\ref{fig1}).
Then, using the dispersion relation $\omega^2 = kg$
we obtain expressions for the real and imaginary parts of $\kappa$ (see Fig.~\ref{fig2}):
\begin{equation}
\begin{split}
& \alpha(\omega) = \Re(\kappa) = \frac{\omega^2\sqrt{2}}{2g}
  \sqrt{1+\sqrt{1+(\Omega/\omega)^6}}\,,  \\
& \beta(\omega) = \Im(\kappa) = \frac{\omega^2\sqrt{2}}{2g}
  \sqrt{-1+\sqrt{1+(\Omega/\omega)^6}}\,.
\end{split}
\end{equation}

Solutions of the equations for the Hertz vector associated with the ring
progressive waves propagating along the surface
of deep fluid with velocity potential (\ref{R7}) are:
\begin{equation}\label{R11}
\begin{split}
& P_z = C(\omega)\frac{\ex^{kz}}{k}\left(2\left(\frac{\omega}
{\Omega}\right)^3\left(F_z+F_y\frac{\partial}{k\partial y}\right)
\left[\left(1-\frac{\alpha}{k}\right)\Lambda_3-\frac{\beta}{k}\Lambda_4
\right]\ex^{(\alpha/k-1)kz}+F_z\Lambda_2\right)\,, \\
& P_{az} = C(\omega)\frac{\ex^{-kz}}{k}\left(2\left(\frac{\omega}
{\Omega}\right)^3\left(F_z+F_y\frac{\partial}{k\partial y}\right)
\left[\left(1-\frac{\alpha}{k}\right)\Lambda_1-\frac{\beta}{k}\Lambda_2
\right]+F_z\Lambda_2\right)\,, \\
& P_y = C(\omega)\frac{\ex^{kz}}{k}F_y\Lambda_2\,, \qquad P_{ay} = -
C(\omega)\frac{\ex^{-kz}}{k}F_y\Lambda_2\,.
\end{split}
\end{equation}
where $C(\omega)=(\omega/g)A(\omega^2/g)$.
\begin{equation}
\begin{split}
& {\Lambda_1(r,t)=\Im(H_0^{(2)}(kr)\exi{\omega t})\,, \qquad
  \Lambda_3(r,t)=\Im(H_0^{(2)}(kr)\exi{(\omega t+\beta z)})\,,  } \\
& {\Lambda_2(r,t)=\Im(\irm H_0^{(2)}(kr)\exi{\omega t})\,, \qquad
  \Lambda_4(r,t)=\Im(\irm H_0^{(2)}(kr)\exi{(\omega t+\beta z)})\,.  }
\end{split}
\end{equation}
Expressions for the components of the electromagnetic field and
electric current density are obtained by simple differentiation (\ref{R11})
by the rule (\ref{R4}).

Using the known asymptotic expansions of the Hankel functions \cite{B6}
write expressions for the electromagnetic quantities in the far zone $r\gg\lambda$.

For the vertical external magnetic field it is true:
\begin{equation}
\begin{split}
& { B_r = F_z G(\omega,r)\left(\ex^{\alpha z}(S(\omega,\psi_1)+
   \cos{\psi_1})-\ex^{kz}\cos{\theta_1}\right)  }\,, \\
& { B_z = F_z G(\omega,r)\left(\ex^{\alpha z}(S(\omega,\psi_0)-
   \cos{\psi_0})+\ex^{kz}\cos{\theta_0}\right)  }\,, \\
& { B_{ar} = F_z G(\omega,r)\ex^{-kz}S(\omega,\theta_1)  }\,, \\
& { B_{az} = F_z G(\omega,r)\ex^{-kz}S(\omega,\theta_0)  }\,,  \\
& { E_{\gamma} = F_z G_0(\omega,r)\left(\ex^{\alpha
z}(S(\omega,\psi_0)-\cos{\psi_0})+\ex^{kz}\cos{\theta_0}\right)  }\,, \\
& { E_{a\gamma} = F_z G_0(\omega,r)\ex^{-kz}S(\omega,\theta_0) }\,,
\\
& { J_{\gamma} = \sigma F_z G_0(\omega,r)\ex^{\alpha
z}(S(\omega,\psi_0)-\cos{\psi_0}) }\,.
\end{split}
\end{equation}
Where the indexes $r$ and $\gamma$ denote, respectively, the radial
and tangential components of the vectors.
\begin{equation}
\begin{split}
& { \theta_m = \omega t -kr+\frac{m\pi}{2}+\frac{\pi}{4}\,, \qquad
\psi_m = \theta_m+\beta z }\,, \\
& { S(\omega,\theta) = 2\left(\frac{\omega}{\Omega}\right)^3
  \left(\frac{\alpha}{k}-1\right)\left(\frac{\beta}{k}\cos{\theta}-
  \sin{\theta}\right)  }\,, \\
& { G(\omega,r) = 2\frac{C(\omega)}{\lambda(\omega)}
   \sqrt{\frac{\lambda(\omega)}{r}}\,,  \quad
   G_0(\omega,r) = \frac{\lambda(\omega)}{\lambda_0(\omega)}
   G(\omega,r)  }\,, \\
& { \lambda(\omega) = 2\pi g/\omega^2 \qquad
           \lambda_0(\omega) = 2\pi c/\omega  }\,.
\end{split}
\end{equation}
The obtained expressions have the following features:
all values decrease exponentially with distance from the surface
of the liquid and are damped by cylindrical law $1/\sqrt{r}$
with the distance from the origin; they are periodically time-dependent,
and with an accuracy of $1/\sqrt{r}$, periodically depend on the
horizontal coordinate. Each value has cylindrical symmetry and has a shape
of progressive wave propagating from the origin.
Induced magnetic field vector lies in a plane passing through the vertical axis.
At each point in space above the liquid surface with the passage of time the
vector rotates, describing a circle.
Beneath the surface this circle is deformed.
Electric field and current have only components directed along the crest of the wave.
Lines of electric current, being confined, form a system of concentric circles.
It should also be noted that due to the effect of self-induction tangential component
of the electric field is different from zero and reaches a maximum
at the frequency $\omega\approx\Omega$.

For the horizontal external magnetic field it is true for the magnetic field in the liquid:
\begin{equation}\label{R14}
\begin{split}
& B_x = \frac{1}{2}F_y G(\omega,r)\left(e^{\alpha z}
  (S(\omega,\psi_0)+\cos{\psi_0})-e^{kz}\cos{\theta_0}\right)
  \sin{2\gamma}\,,   \\
& B_y = \frac{1}{2}F_y G(\omega,r)\left(e^{\alpha z}
  (S(\omega,\psi_0)+\cos{\psi_0})-e^{kz}\cos{\theta_0}\right)
  (1-\cos{2\gamma})\,, \\
& B_z = F_y G(\omega,r)\left(e^{\alpha z}
  (S(\omega,\psi_1)-\cos{\psi_1})+e^{kz}\cos{\theta_1}\right)
  \sin{\gamma}\,, \\
\end{split}
\end{equation}
For the magnetic field in the air:
\begin{equation}
\begin{split}
& B_{ax} = \frac{1}{2}F_y G(\omega,r)e^{-kz}
  S(\omega,\theta_0)\sin{2\gamma}\,, \\
& B_{ay} = \frac{1}{2}F_y G(\omega,r)e^{-kz}
  S(\omega,\theta_0)(1-\cos{2\gamma})\,, \\
& B_{az} = F_y G(\omega,r)e^{-kz}S(\omega,\theta_1)
  \sin{\gamma}\,,  \\
\end{split}
\end{equation}
For the electric current an electric field in the liquid:
\begin{equation}
\begin{split}
& J_x = \sigma\frac{1}{2}F_y G_0(\omega,r)e^{\alpha z}
  (S(\omega,\psi_1)-\cos{\psi_1})(1-\cos{2\gamma})\,, \\
& J_y = -\sigma\frac{1}{2}F_y G_0(\omega,r)e^{\alpha z}
  (S(\omega,\psi_1)-\cos{\psi_1})\sin{2\gamma})\,, \\
& E_x = \frac{1}{2}F_y G_0(\omega,r)\left( e^{\alpha z}
  (S(\omega,\psi_1)-\cos{\psi_1})(1-\cos{2\gamma})+
  2 e^{kz}\cos{\theta_1} \right)\,, \\
& E_y = -\frac{1}{2}F_y G_0(\omega,r)e^{\alpha z}
  (S(\omega,\psi_1)-\cos{\psi_1})\sin{2\gamma})\,, \\
& E_z = F_y G_0(\omega,r)e^{kz}\sin{\theta_1}
  \cos{\gamma}\,, \\
\end{split}
\end{equation}
For the electric field in the air and surface electric charge:
\begin{equation}\label{R14a}
\begin{split}
& E_{ax} = \frac{1}{2}F_y G_0(\omega,r)e^{-kz}\left(
  (S(\omega,\theta_1)-\cos{\theta_1})(1-\cos{2\gamma})+
  2\cos{\theta_1} \right)\,, \\
& E_{ay} = -\frac{1}{2}F_y G_0(\omega,r)e^{-kz}
  (S(\omega,\theta_1)-\cos{\theta_1})\sin{2\gamma})\,, \\
& E_{az} = -F_y G_0(\omega,r)e^{-kz}\sin{\theta_1}
  \cos{\gamma}\,, \\
& q = -\frac{1}{2\pi}F_y G_0(\omega,r)\sin{\theta_1}
  \cos{\gamma}\,.
\end{split}
\end{equation}
As follows from the expressions (\ref{R14}--\ref{R14a}) for the case of
the horizontal external magnetic field all three components
of the induced electric and magnetic fields in the air and
liquid are non vanishing.
And their values depend strongly on the azimuthal angle $\gamma$.

The electric current lines form closed configurations, symmetric
with respect to the origin and axis $x$ and $y$.
Condition of electric current impermeability through the surface
leads to a surface charge $q$, and hence to the electrostatic
field $E_{az}$, $E_z$.

\section{Electromagnetic fields induced by nonstationary\\ \medskip ring waves}

In this section we solve the problem of obtaining analytical expressions
for the electromagnetic fields induced by waves of Cauchy-Poisson in a
constant magnetic field.

In nature these dispersive nonstationary waves formed by decay of the
initial localized disturbance may arise for example from a stone thrown
into the water.

In some cases, to the Cauchy-Poisson problem the description of tsunami
waves generation in the ocean is reduced~\cite{B1}.
In this regard the study of electromagnetic fields, produced by such surface
wave disturbances of conducting fluid in a constant external magnetic field
has a practical application in the development of tsunamis early detection systems.
It is particularly important
that the electromagnetic field induced by the tsunami waves can be registered
before the waves themselves, they are a kind of electromagnetic tsunami precursors.

The problem is considered in the approximation of the deep ocean,
a uniform with respect on electrical conductivity.
Coordinate system is chosen so that the direction of the axis $y$
coincides with the direction of the horizontal component of the geomagnetic
field $\vec F$. Axis $z$ is directed upward.
Quantities relating to water are taken without an index,
and the values in the air with an index $a$.

The problem will be solved by writing Maxwell's equations through the
Hertz magnetic vector potential:
\begin{equation}\label{S1}
\nu_m\Delta\vec P - {\partial_t}\vec P = \phi {\vec F}\,.
\end{equation}
Here the magnetic Hertz vector $\vec P$ has only $y$ and $z$ components;
$\nu_m = {c^{2}/4\pi\sigma}$ is magnetic viscosity, $c$ is the speed of light,
$\sigma$ is electrical conductivity of the fluid, $\Delta$ is Laplace operator,
$\phi$ is velocity potential of fluid. In the air the equation (\ref{S1})
is transformed into the Laplace equation.

The components of the electromagnetic field and the electric current density
can be found through the Hertz vector from the following expressions:
\begin{equation}\label{S2}
\begin{split}
& {\vec B}=\rot\rot{\vec P} = \grad\Div{\vec P}-\Delta{\vec P}\,,  \\
& {\vec E}=-\frac{1}{c}\rot{\partial_t{\vec P}}\,, \\
& {\vec J}=-\frac{\sigma}{c}\rot\left({\partial_t{\vec P}}
+\phi{\vec F}\right) = -\frac{c}{4\pi}\rot\Delta{\vec P}\,, \\
& {\vec v} = - \grad\phi\,,
\end{split}
\end{equation}
where ${\vec B}$ is the magnetic induction vector,
${\vec E}$ is the electric field vector,
${\vec J}$ is the electric current density vector,
${\vec v}$ is fluid velocity field.

Equation (\ref {S1}) is solved for each of the media by using the
following boundary conditions at z = 0:
\begin{equation}\label{S3}
P_z=P_{az}\,, \quad {\partial_z{P_y}} =
 {\partial_z{P_{ay}}}\,,  \quad
{\partial_z{P_z}}+{\partial_y{P_y}}=
 {\partial_z{P_{az}}}+{\partial_y{P_{ay}}}\,, \quad
\Delta P_y = \Delta P_{ay}\,.
\end{equation}
Due to the non-stationarity of the Cauchy-Poisson we use
the method of the Laplace transform.
In accordance with this method associate with the Hertz vector potential
and with the speed potential their integral images: $\Ls(\vec P(t))=\vec u(p)$;
$\Ls(\phi(t))=s(p)$. Equation (\ref {S1}) is also subjected to the
Laplace transform with zero initial conditions for the $\vec P(t)$.

As a result we have a system of equations for $\vec u(p)$.
By solving it with the boundary conditions (\ref{S3}) we find this function
and by returning from the images to the originals, we obtain the solution for $\vec P (t)$.
\begin{equation}\label{S4}
\nu_m \Delta{\vec u}-p\vec u= s\vec F\,, \qquad \Delta\vec u_a = 0\,.
\end{equation}
First obtain solutions for the elementary potential of the following form:
\begin{equation}\label{S5}
\begin{split}
& \phi(t) = 0 \mbox{ для } t<0 \mbox{ и } \\
& \phi(t) = R(x,y)e^{kz}\sin{\omega t} \mbox{ для } t \ge 0\,.
\end{split}
\end{equation}
Its image $s(p) = Re^{kz}(\omega/(\omega^2+p^2))\,.$
Here $k$ is wavenumber, $\omega$ is frequency,
$R(x,y)$ is function of the horizontal coordinates of
general form satisfying the Helmholtz equation $R_{xx}+R_{yy}+k^2R = 0$.
\begin{equation}\label{S6}
\begin{split}
& u_y = -G(p)F_y e^{kz}R(x,y)\,, \\
& u_{ay} = G(p)F_y e^{-kz}R(x,y)\,, \\
& u_z = M(p)e^{\eta(p)z}\left(F_z+F_y\frac{\partial}{k\partial y}
  \right)R(x,y)+  \\
& \qquad + G(p)\left(F_z(e^{\eta(p)z}-e^{kz})+F_y\frac{\partial}
  {k\partial y}\right)R(x,y)\,, \\
& u_{az} = M(p)e^{-kz}\left(F_z+F_y\frac{\partial}{k\partial y}
  \right)R(x,y)+ \\
& \qquad + G(p)e^{-kz}F_y\frac{\partial}{k\partial y}R(x,y)\,.
\end{split}
\end{equation}
Here we denote $\eta(p)=\sqrt{k^2+\nu_m^{-1}p}\,.$
\begin{equation}
\begin{split}
& { G(p) = \frac{1}{p}\frac{\omega}{\omega^2+p^2} }\,, \\
& { M(p) = \frac{1}{\Omega^2}\frac{2\Omega}{\Omega^2+(p/\alpha)^2}
  \left(\frac{\alpha}{p}\right)^2\left[\left(\sqrt{1+(p/\alpha)^2}-1\right)
  -\frac{1}{2}(p/\alpha)\right]  }\,,
\end{split}
\end{equation}
where $\alpha=\nu_m k^2$ and $\Omega = \omega/\alpha\,.$

From the obtained images, one can restore the original functions.
Confine ourselves to the fields in the air and restore original functions
from $G(p)$ and $M(p)$:
\begin{equation}\label{S7}
\begin{split}
& { \Ls^{-1}(G(p))= f_1(t) = \frac{1}{\omega}(1-\cos(\omega t)) }\,, \\
& { \Ls^{-1}(M(p))= f_2(t) = f(\alpha t) }\,, \\
& { f(t) = \frac{2}{\alpha\Omega}\left(\frac{\sin\Omega t}{\Omega}+
  \frac{\cos\Omega t}{2}+\left[\left(t+\frac{1}{2}\right)(\erf(\sqrt{t})-1)
  +\sqrt{\frac{t}{\pi}}e^{-t}\right] -\psi(t)\right) }\,,\\
& { \psi(t) = \frac{1}{\Omega}\Im\left(\sqrt{1+i\Omega}e^{i\Omega
t}\erf\left(\sqrt{1+i\Omega}\sqrt{t}\right)\right) = } \\
& \qquad = \frac{1}{\Omega}\frac{e^{-t}}{\sqrt{\pi t}}\Im\left(
 \Phi(1,1/2,(1+i\Omega)t)\right)\,.
\end{split}
\end{equation}
Here $\erf(x)$ is probability integral and $\Phi(a,c,z)$ is degenerate
Kummer hypergeometric function.

For the obtained function $f(t)$, we can write various asymptotic estimates.
At large times $t\gg 1$ we have:
\begin{equation}
\begin{split}
& f(t) = \frac{2}{\alpha\Omega^2}\Im\left((1+i\Omega/2)-
\sqrt{1+i\Omega}\right)
+\frac{e^{-t}}{2t\sqrt{\pi t}}\left(1+\frac{1}
{1+\Omega^2}\right)+O\left(\frac{e^{-t}}{t^{5/2}} \right)\,, \\
& f(t) = \frac{\Omega}{\alpha}\left(\frac{\cos{\Omega t}}{8}-
  \frac{\sin{\Omega t}}{4\Omega}-
\frac{e^{-t}}{t\sqrt{\pi t}}
+O\left(\frac{e^{-t}}{t^{5/2}} \right)
   \right) \mbox{ для } \Omega \rightarrow 0\,.
\end{split}
\end{equation}
Let us consider the dimensionless parameter $\Omega$:
\begin{equation}
\Omega=\omega/\alpha=(kg)^{1/2}/(\nu_m k^2)=(l/l_m)^{3/2}\,, \qquad
l_m=(\nu_m^2/g)^{1/3}\,.
\end{equation}
\begin{figure}[tbh]
\centering{%
\includegraphics[width=0.65\textwidth,angle=-90]{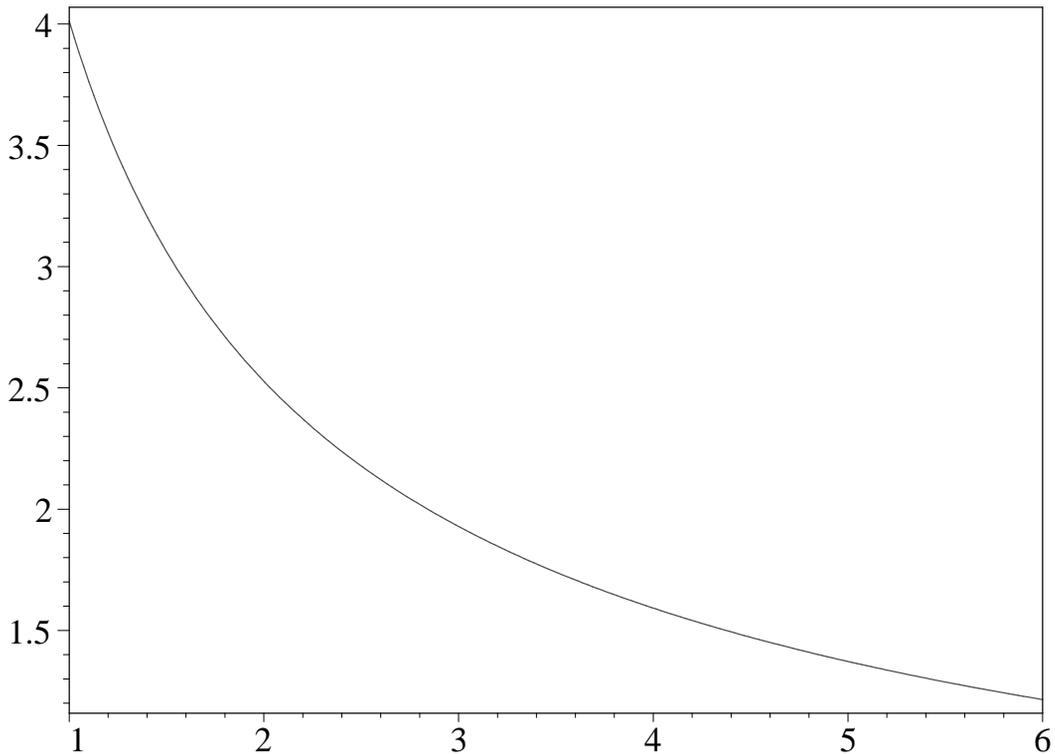}}
\caption{Dependance of $l_m$ (km) from the conductivity of the sea water $\sigma$ (Sm/m).}
\label{fig3}
\end{figure}

It turns out that the value of $\Omega$ is determined by the ratio of the
characteristic size of the perturbation $l$ to electromagnetic
length $l_m$, which depends on the electrical conductivity
of sea water and sets in our case a natural length scale (see Fig.~\ref{fig3}).
Qualitative behavior of solutions depends on relative sizes of wave
disturbances with respect to this natural scale.
With the decrease of the characteristic size
of the perturbation $\Omega$ is also reduced.
If $\Omega\ll 1$, then for all times up to terms of second order
in $\Omega$, we have:
\begin{equation}
\begin{split}
& \alpha f(t) = -\frac{1}{4}\sin(\Omega t)+
\Omega\left(f_0(t)+\frac{1}{8}\cos(\Omega t)\right)\,, \\
& f_0(t) = \left(\frac{1}{8}-\frac{t}{4}-\frac{t^2}{2}-
\frac{t^3}{3} \right)(\erf(\sqrt{t})-1)+
\left(\frac{t^2}{3}+\frac{t}{3}-\frac{1}{4}\right)
\sqrt{\frac{t}{\pi}}\exp(-t)\,.
\end{split}
\end{equation}
In addition it is true the following decomposition at small times:
\begin{equation}
\begin{split}
& f_0(t) = \left(\frac{1}{8}-\frac{t}{4}-\frac{t^2}{2}-
\frac{t^3}{3} \right) - \\
& \qquad - 16\sqrt{\frac{t}{\pi}}
\sum_{k=2}^{\infty}\frac{(-1)^k t^k}{(k-2)!}
\left(\frac{1}{(2k+1)(2k-1)(2k-3)(2k-5)}\right)\,, \\
& f_0(t) = \frac{\Omega}{\alpha}\left[-\left(\frac{t^2}{2}+
\frac{t^3}{3}\right)+\frac{16}{15}\sqrt{\frac{t}{\pi}}
(t^2+\frac{1}{7}t^3)+O(t^{9/2})\right]\,.
\end{split}
\end{equation}
Finally we write the Hertz vector excited by elementary
potential (\ref{S5}) in the air:
\begin{equation}\label{S11}
\begin{split}
& { P_{ay} = f_1(t)F_y e^{-kz}R(x,y)  }\,,  \\
& { P_{az} = f_2(t)e^{-kz}\left(F_z+F_y\frac{\partial}{k\partial y}
  \right)R(x,y)+ } \\
& \qquad + f_1(t)e^{-kz}F_y\frac{\partial}{k\partial y}R(x,y)\,.
\end{split}
\end{equation}
Electromagnetic fields induced by the decaying initial disturbance
of the liquid surface will match the speed potential \cite{B3}:
\begin{equation}\label{S12}
\phi = \int\limits_0^\infty \sqrt{kg}e^{kz}\sin{\sqrt{kg}t}
 J_0(kr)\int\limits_0^\infty \alpha J_0(k\alpha) N(\alpha)\,d\alpha\,dk\,.
\end{equation}
Where $J(kr)$ is Bessel function, $N(r)$ is initial radially-symmetric
shape of the liquid surface.

To find the components of the Hertz vector corresponding
to such a speed potential, it is necessary in the expressions
(\ref{S11}) for $R(x, y)$ put
$R(x,y)=\omega J_0(kr)\int_0^{\infty}\alpha J_0(k\alpha)N(\alpha)\,d\alpha$
and integrate it over $k$ in the semi-infinite range:
\begin{equation}\label{S13}
\begin{split}
& P_{ay} = F_y\int_0^{\infty}f_1(t)\sqrt{kg}e^{-kz}J_0(kr)\,dk
\int_0^{\infty}\alpha J_0(k\alpha)N(\alpha)\,d\alpha\,, \\
& P_{az}=-F_y\sin(\vartheta)\int_0^{\infty}f_1(t)\sqrt{kg}e^{-kz}
J_1(kr)\,dk\int_0^{\infty}\alpha J_0(k\alpha)N(\alpha)\,d\alpha + \\
&  + \int_0^{\infty}f_2(t)\sqrt{kg}e^{-kz}(F_z J_0(kr)-F_y
 J_1(kr)sin(\vartheta))\,dk
\int_0^{\infty}\alpha J_0(k\alpha)N(\alpha)\,d\alpha\,.
\end{split}
\end{equation}
Here $\vartheta$ is the polar angle.

These are the final general solutions of the problem
determining the electromagnetic fields in the air induced
by decay of radially-symmetric initial perturbations of the
surface of a conducting liquid in a constant external magnetic
field at low magnetic Reynolds number $Re_m$.
Components of electromagnetic quantities can be found by the formulas (\ref{S2}).
If you want to get electromagnetic fields induced by initial pulse pressure,
then the solutions (\ref{S13}), differentiated by time, will give us
the desired for the initial pressure $P(r)=(\rho/g)N(r)$,
where $\rho$ is liquid density, $g$ is the acceleration of gravity.

\subsection*{Behavior of the solutions on the axis $r=0$}

\begin{figure}[tbh]
\centering{%
\includegraphics[width=0.65\textwidth,angle=-90]{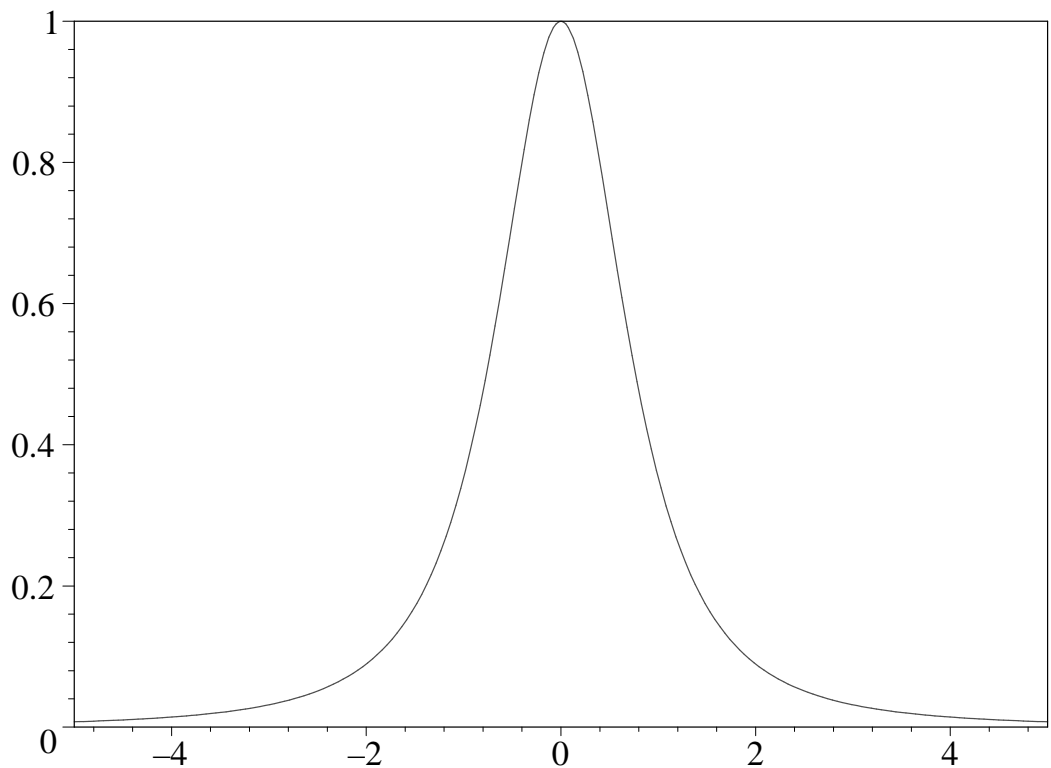}}
\caption{Initial form of the sea surface $N(r)$ for $A=1$ and $a=1$.}
\label{fig4}
\end{figure}

\begin{figure}[tbh]
\centering{%
\includegraphics[width=0.65\textwidth,angle=-90]{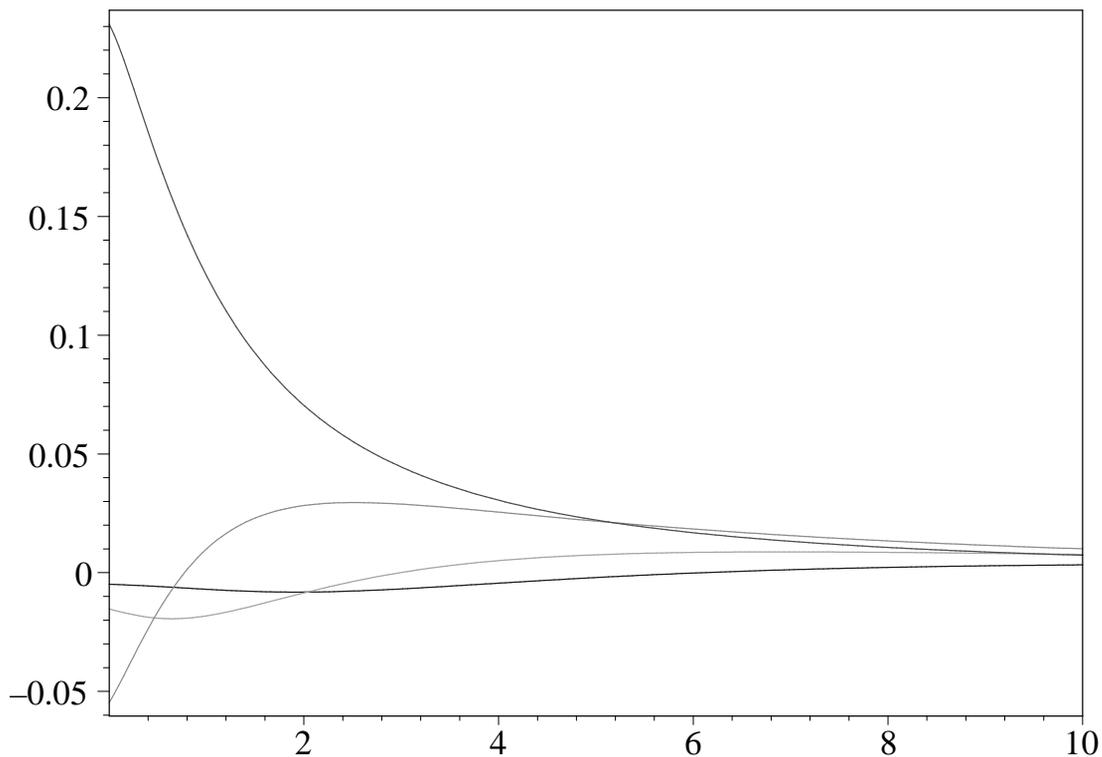}}
\caption{Function $\frac{\tau}{(1+\mu)^2}\Phi(2,3/2,-\tau^2/(1+\mu))$
from $\mu$ at $\tau=1,2,3,4$.}
\label{fig5}
\end{figure}

We now consider various asymptotic behavior of the obtained solutions.
Assume for definiteness that the initial liquid surface shape is (see Fig.~\ref{fig4}):
\begin{equation}\label{S14}
N(r) = A [1+(r/a)^2]^{-3/2}\,.
\end{equation}
Suppose also that the size of this perturbation is small enough
so that for almost all values of the wave number, forming it true
$\Omega\ll 1$, in fact it is necessary $a\ll l\approx 1.5$ km.
Let $r = 0$ and write solutions for the nonzero components of
the electromagnetic field on this axis (see Fig.~\ref{fig5}).
\begin{equation}\label{S15}
\begin{split}
& B_{az}(z,t) = -\frac{1}{2}F_z Re_m\frac{\tau}{(1+\mu)^2}
\Phi(2,3/2,-\tau^2/(1+\mu))\,, \\
& B_{ay}(z,t) = -\frac{1}{2}F_y Re_m\frac{\tau}{(1+\mu)^2}
\Phi(2,3/2,-\tau^2/(1+\mu))\,, \\
& E_{ax}(z,t) = -F_y L_m\frac{\tau}{(1+\mu)^3}
\Phi(3,3/2,-\tau^2/(1+\mu))\,, \\
& \zeta(0,t)=A\Phi(2,1/2,-\tau^2)\,.
\end{split}
\end{equation}
Here $\mu=z/a$ is dimensionless height; $\tau=(g/4a)^2t$
is dimensionless time. We have also introduced the parameters
$Re_m=A\sqrt{ga}/\nu_m$ and $L_m=2(A/a)(\sqrt{ga}/c)$.
It is interesting to see how the formulas (\ref{S15})
behave at various values of the parameters $\mu$ and $\tau$.
For small times, when $\tau^2/(1+\mu)\ll 1$, we obtain:
\begin{equation}\label{S16}
\begin{split}
& B_{az}(z,t) = -\frac{1}{2}F_z Re_m\frac{\tau}{(1+\mu)^2} \\
& B_{ay}(z,t) = -\frac{1}{2}F_y Re_m\frac{\tau}{(1+\mu)^2} \\
& E_{ax}(z,t) = -F_y L_m\frac{\tau}{(1+\mu)^3}
\end{split}
\end{equation}
It appears that in this case all the components increase linearly
with time, and decrease in space by the power law.
Even more interesting behavior is found after a long time near
the surface.
Use known asymptotic of Kummer function
$\Phi(a,c,x) = (\Gamma(c)/\Gamma(c-a))(-x)^{-a}(1+O(|x|^{-1}))$
and for large argument we obtain for $\tau^2/(1+\mu) \gg 1$:
\begin{equation}\label{S17}
\begin{split}
& B_{az}(z,t) = \frac{1}{8}F_z Re_m \tau^{-3} \\
& B_{ay}(z,t) = \frac{1}{16}F_y Re_m \tau^{-3} \\
& E_{ax}(z,t) = -\frac{3}{8}F_y L_m \tau^{-5}
\end{split}
\end{equation}
That is, the field components near the origin does not depend
on the vertical coordinate, and decrease with time according
to a power law. In addition, there are situations where at one
point in time the field at the surface is zero, but with increasing
altitude, it appears at a certain height reaches a maximum and then
begins to gradually subside.

\subsection*{Behavior of the solutions for large $r$ and $t$}

\begin{figure}[tbh]
\centering{%
\includegraphics[width=0.65\textwidth,angle=-90]{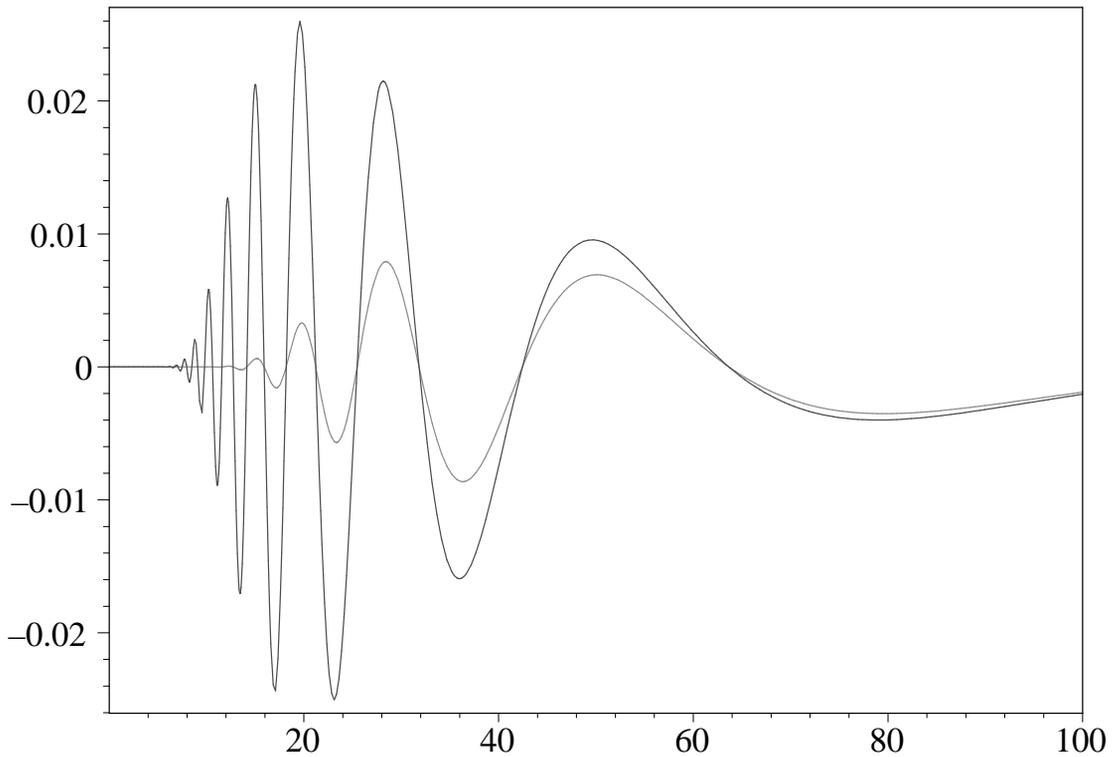}}
\caption{Functions $I_0(20,0,\chi)$ and $I_0(20,2,\chi)$ from $\chi$ for $\tau=20$ and $\mu=0,2$.}
\label{fig6}
\end{figure}

Find out how the components of the field behave in the space,
if the time elapsed since the dissolution of the initial disturbance
is relatively large. And because, as was explained, electric and magnetic
fields are qualitatively similar in their behavior, we restrict ourselves
to the magnetic field.
Study formulas (\ref{S13}) with the initial perturbation (\ref{S14})
by stationary phase method \cite{B4,B5,B6} and get the following results:
\begin{equation}\label{S18}
\begin{split}
& B_{ax} =-\frac{1}{8}Re_m (2F_z I_1\cos(\vartheta)
 + F_y I_0\sin(2\vartheta))\,, \\
& B_{ay} =-\frac{1}{8}Re_m (2F_z I_1\sin(\vartheta)
 + F_y I_0(1-\cos(2\vartheta) ))\,, \\
& B_{az} =-\frac{1}{4}Re_m (F_z I_0 + F_y I_1\sin(\vartheta))\,.
\end{split}
\end{equation}
Here functions $I_0$ and $I_1$ are (for $\chi=r/a$, $\mu\ll\chi$):
\begin{equation}\label{S19}
\begin{split}
& I_0 = \sqrt{2}\frac{\tau}{\chi^2}\exp(-\frac{\tau^2}{\chi^2}[1+\mu])
\sin(\frac{\tau^2}{\chi}), \\
& I_1 = \sqrt{2}\frac{\tau}{\chi^2}\exp(-\frac{\tau^2}{\chi^2}[1+\mu])
\cos(\frac{\tau^2}{\chi})\,.
\end{split}
\end{equation}
As expected, the components of the magnetic field form in space a package
of oscillation which propagates from the origin with speed $v = (g(a+z)/4)^{1/2}$
at each height (see Fig.~\ref{fig6}).
Speed of the package increases with increase of altitude.
This effect can be explained by the fact that in infinitely deep sea with
increase of the length of the harmonic wave its phase velocity unlimitedly increases.
However, since there is an exponential attenuation of the induced fields
from the individual harmonics with the growth of the height,
the greater attenuation for the shorter wavelength, then for the high altitude
the shortwave components are filtered out, and the remaining faster long-wave
components make the main contribution to the variation of the field.

\end{document}